\documentclass[cernpreprint, catlasdraft=true, texlive=2023, UKenglish, orcidlogo, texmf, orcidlogo]{atlasdoc}
 
\usepackage{atlaspackage}
\usepackage{atlasbiblatex}
 
\usepackage{atlasphysics}

\graphicspath{{logos/}{figures/}}
 
\usepackage{pdflscape}
\usepackage{ANA-HIGG-2021-05-PAPER-defs}

\AtlasTitle{Combination of searches for invisible decays of the Higgs boson using 139~{\mbox{fb\(^{-1}\)}} of proton-proton collision data at $\sqrt{s} = 13~$TeV collected with the ATLAS experiment}

\AtlasAbstract{
Many extensions of the Standard Model predict the production of dark matter particles at the
LHC. Sufficiently light dark matter particles may be produced in decays of the Higgs boson
that would appear invisible to the detector. This Letter presents a statistical combination
of searches for $H\to\text{invisible}$~decays where multiple production modes of the Standard Model Higgs boson are considered.
These searches are performed with the ATLAS detector using 139 fb$^{-1}$
of proton--proton collisions at a centre--of--mass energy of $\ensuremath{\sqrt{s} =\SI{13}{TeV}}\xspace$ at the LHC. In combination
with the results at $\ensuremath{\sqrt{s} =\SI{7}{TeV}}\xspace$ and $\ensuremath{\SI{8}{TeV}}\xspace$, an upper limit on the $H\to\text{invisible}$~branching ratio
of 0.107 (0.077)
at the 95\% confidence level is observed (expected). These results are also
interpreted in the context of models where the 125~GeV Higgs boson acts as a portal to dark
matter, and limits are set on the scattering cross-section of weakly interacting massive particles
and nucleons.}
 
\AtlasRefCode{HIGG-2021-05}
 
\PreprintIdNumber{CERN-EP-2022-289}

\AtlasJournal{Phys.\ Lett.\ B.}
\AtlasJournalRef{Phys. Lett. B 842 (2023) 137963}
\AtlasDOI{10.1016/j.physletb.2023.137963}

\hypersetup{pdftitle={ATLAS document},pdfauthor={The ATLAS Collaboration}}
 
\begin{document}
 
\maketitle
 
\tableofcontents

\section{Introduction}
 
Compelling astrophysical evidence suggests that dark matter (DM) comprises most of the matter in
the universe~\cite{Zwicky:1933gu,DMreview,Komatsu:2010fb,Planck:2015fie}. However, its nature is still unknown and poses one of the central questions in modern physics.
A possible candidate for DM is a massive, stable and electrically neutral particle $\chi$, interacting weakly with the known particles of the Standard Model (SM).

Several theoretical frameworks predict the production of DM particles in proton--proton collisions at the Large Hadron
Collider (LHC)~\cite{Abercrombie:2015wmb,Feng_2010,Boveia_2018}. In a wide class of those models,
the 125~\GeV\ Higgs boson~\cite{HIGG-2012-27,CMS-HIG-12-028} acts as a portal between a dark sector and the SM sector, either through Yukawa-type couplings to fermionic DM, or other mechanisms~\cite{Antoniadis:2004se,ArkaniHamed:1998vp,Datta:2004jg,Kanemura:2010sh,Djouadi:2011aa,Djouadi:2012zc,Shrock:1982kd,Choudhury:1993hv,Eboli:2000ze,Davoudiasl:2004aj,Godbole:2003it,Ghosh:2012ep,Belanger:2013kya,Curtin:2013fra}. If kinematically allowed, pairs of DM particles can then be produced via the decay of the Higgs boson.
The DM particles would traverse the detector without interacting and are inferred indirectly through the presence of
missing transverse momentum (\met)\footnote{ATLAS uses a right-handed coordinate
system with its origin at the nominal interaction point (IP) in the
centre of the detector and the $z$-axis along the beam pipe. The
$x$-axis points to the centre of the LHC ring, and the
$y$-axis points upward. Cylindrical coordinates $(r,\phi)$ are used
in the transverse plane, $\phi$ being the azimuthal angle around the
$z$-axis. The pseudorapidity is defined in terms of the polar angle
$\theta$ as $\eta=-\ln\tan(\theta/2)$. The distance between two
objects in $\eta$--$\phi$ space is $\Delta R = \sqrt{(\Delta \eta)^2
+ (\Delta \phi)^2}$. Transverse momentum is defined by $\pt =  p\sin{\theta}$.} in the interaction.
This decay channel is therefore called ``invisible.'' In the SM, the branching fraction to invisible final states is about 0.1\%~\cite{deFlorian:2016spz}
arising from the decay of the Higgs boson via \(ZZ^* \rightarrow 4\nu\).
 
Direct searches for invisible decays of the Higgs boson were carried out with the ATLAS detector~\cite{PERF-2007-01,ATLAS-TDR-2010-19} during Run~1 of the LHC, using up to 4.7~\ifb\ of $pp$ collision data at a centre-of-mass energy of $\sqrt{s} = 7~\TeV$ and up to 20.3~\ifb\ at 8~TeV. Various event topologies were considered: vector boson fusion (VBF)~\cite{HIGG-2013-16}, production in association with a $Z$ boson (\(ZH\)) that decays into a pair of electrons or muons~\cite{HIGG-2013-03}, and with a $W$ or $Z$ boson that decays into hadrons~\cite{HIGG-2014-07}.
A statistical combination of these ATLAS searches resulted in an observed (expected) upper limit at the 95\% confidence level~(CL) on the invisible Higgs boson branching ratio of $\brinv~<~0.25~(0.27^{+0.10}_{-0.08})$~\cite{HIGG-2015-03}.
These searches were expanded with up to 36~\ifb\ of Run~2 data and their combination, including Run~1 results, yielding an upper limit of $\brinv~<~0.26~(0.17^{+0.07}_{-0.05})$ at the 95\% CL~\cite{HIGG-2018-54}. A combination from the CMS experiment using a similar dataset reported an observed (expected) upper limit of 0.19 (0.15)~\cite{CMS-HIG-17-023}.
 
More recently, new direct searches for invisible decays of the Higgs boson using the full Run 2 data of up to 139~\ifb\ were performed, covering most of the Higgs boson production modes, by ATLAS~\cite{EXOT-2020-11,EXOT-2021-17,ATLAS-CONF-2022-007,HIGG-2018-26,EXOT-2018-06} and CMS~\cite{CMS-EXO-19-003,CMS-HIG-20-003,CMS-EXO-20-004}. In both experiments the VBF final state is the most sensitive channel resulting in an upper limit of 0.145 (0.103) for ATLAS and 0.18 (0.12) for CMS.
 
A partial combination of the ATLAS VBF and \(ZH\) searches together with the analyses targetting visible decays of the Higgs boson was carried out~\cite{ATLAS:2022vkf} and its results reduce the observed (expected) upper limit on \brinv to $0.13$ ($0.08$). Such a combination considers the impact of \brinv on the Higgs boson total decay width and simultaneously determines \brinv, together with the coupling of the Higgs boson to all the SM particles as well as a potential contribution to undetected Higgs boson decays not generating missing transverse energy. The approach relies on a different set of assumptions to what is used in this letter.

This letter presents the statistical combination of all ATLAS direct searches for invisible decays of the Higgs boson using the full Run2 dataset. This includes the gluon--gluon fusion, VBF, \(ZH\) and \(ttH\) production modes, represented in Figure~\ref{fig:diagrams} and assumes the production cross-sections of the Higgs boson does not deviate from the SM predictions~\cite{deFlorian:2016spz, Lidnert:2018kmw, Djouadi:2018kms, Bonetti:2018ukf, Dulat:2018rbf, Harlander:2018yio, Cacciari:2018dksz}.
In addition, a statistical combination with the combined Run~1 result~\cite{HIGG-2015-03} from ATLAS is included,
yielding the most sensitive direct constraint to invisible Higgs boson decays in ATLAS.
 
\begin{figure}[htbp!]
\centering
\subfloat[VBF topology]{\includegraphics[height=0.23\linewidth]{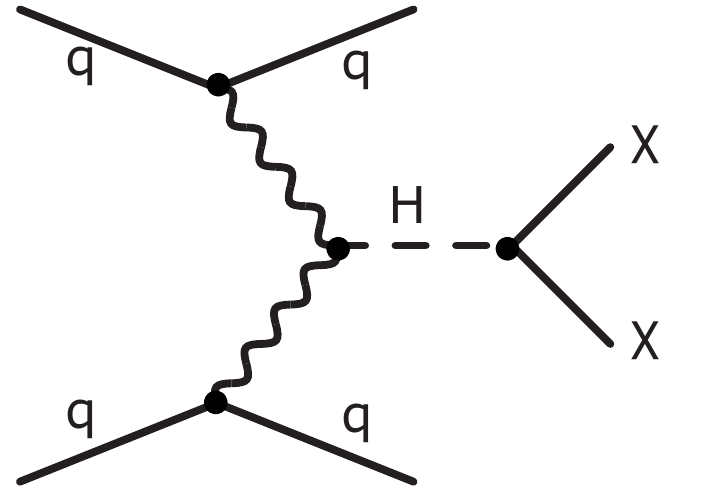}}
\quad
\subfloat[$Z+H$ topology]{\includegraphics[height=0.22\linewidth]{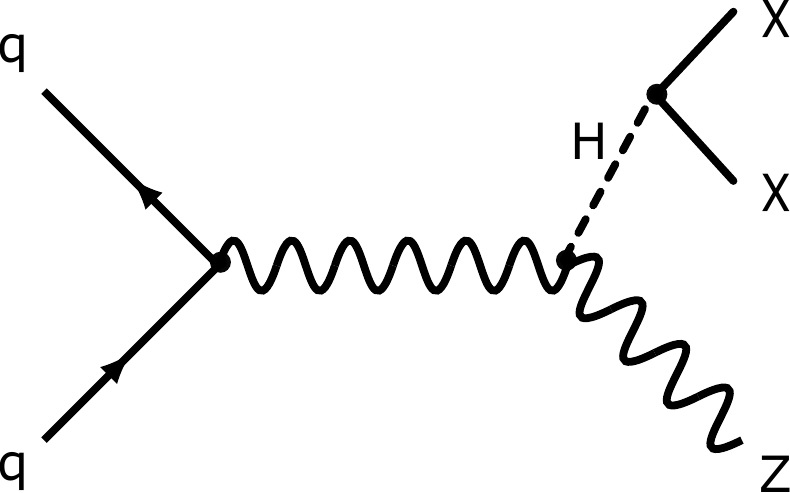}}
\qquad
\subfloat[$\ttbar H$ topology]{ \includegraphics[height=0.23\linewidth]{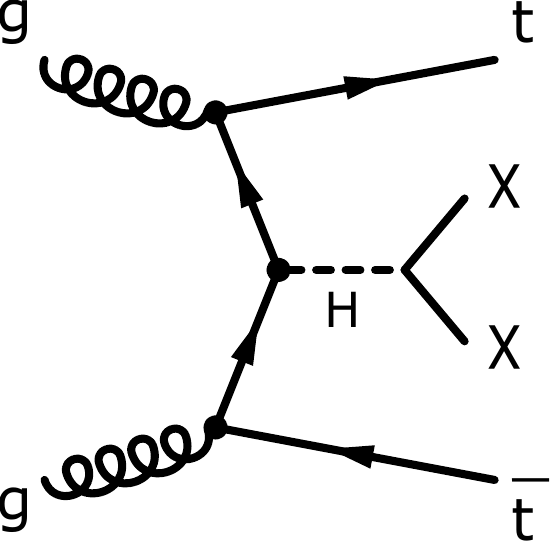}}
\\
\subfloat[VBF + photon topology]{\includegraphics[height=0.23\linewidth]{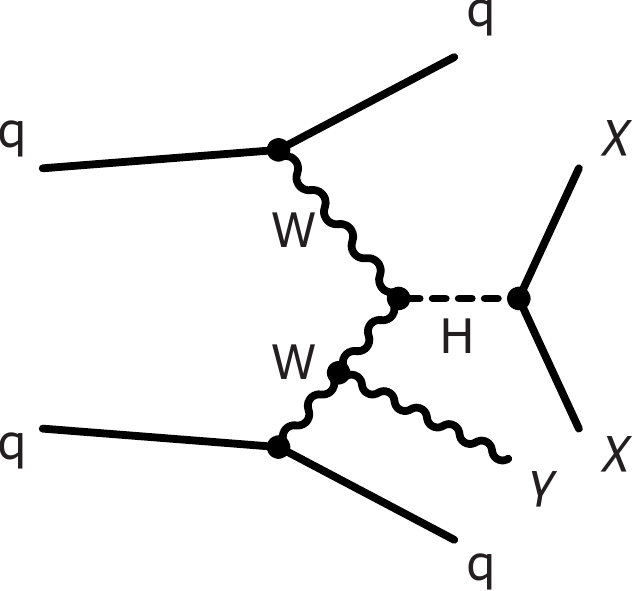}}
\qquad
\subfloat[$H$+jet topology]{\includegraphics[height=0.23\linewidth]{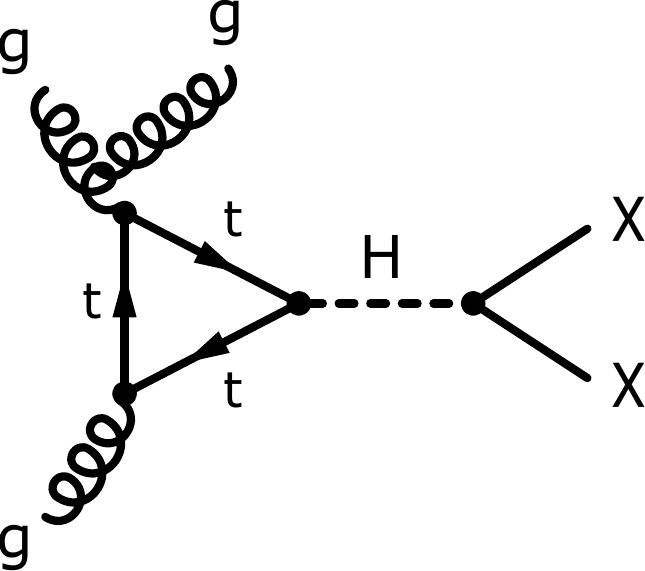}}
\caption{Diagrams illustrating the Higgs boson production mode targetted for the Run~2 searches.}
\label{fig:diagrams}
\end{figure}


\section{Combination inputs}

The inputs to the combination for the Run 2 result consist of searches for invisible decays of the Higgs boson, with the following production modes:
\begin{itemize}
\item VBF topology (\vbf)~\cite{EXOT-2020-11}
\item associated production with a \Zboson boson decaying into electrons or muons (\zll)~\cite{HIGG-2018-26}
\item associated production with a \ttbar pair, using all top-quark decay modes except those with hadronically decaying $\tau$--leptons (\ttH) ~\cite{ATLAS:2022ygn}
\item VBF topology in association with an emitted photon (\vbfGam)~\cite{EXOT-2021-17}
\item gluon--gluon fusion, in association with a high \pT jet (\Mj)~\cite{EXOT-2018-06}
\end{itemize}
all of which use the full data sample, corresponding to an integrated luminosity of 139 fb$^{-1}$.
 
These analyses target different production modes of the Higgs boson and so their event selection criteria are made to be largely orthogonal by using different requirements on lepton, photon, jet and b-tagged jet multiplicity.
The level of residual non-orthogonality was evaluated by considering both the data events and signal samples for all the Higgs boson production modes.
The largest set of shared events is between the \Mj and \vbf searches, which  select  events with large missing transverse energy, no reconstructed leptons, and multiple jets in the final state. The number of overlapping events corresponds to 0.2\% (1.5\%) of the total data (expected signal Monte Carlo (MC) samples) events selected by the \Mj analysis. The impact of the overlap on the final combined result is negligible, altering the upper limit on \brinv by less than 0.001.
A brief overview of the Run~2 analyses and the inputs to the Run~1 combined result~\cite{HIGG-2015-03} is given below.
 
\subsection{\vbf search}
 
In the VBF production mode, the \hinv signal is characterised by two jets with a large separation in pseudorapidity and missing transverse momentum arising from the invisible decays of the Higgs boson. The analysis targetting this signature selects events collected with a trigger selection based on the presence of \met.
Events are further selected if their two jets with the highest \pT fulfill the VBF topology requirements: lying in opposite longitudinal hemispheres, being well separated in $\eta$, and not back-to-back in the transverse plane.
In order to reduce the contribution from $W$, $Z$+jets and \ttbar production, and to ensure orthogonality with the other analyses, events containing leptons or photon candidates, or two or more jets identified as b-tagged jets~\cite{FTAG-2018-01} are vetoed.

In this signature, the dominant background sources are $Z(\rightarrow\nu\nu)+\text{jets}$ and $W(\rightarrow\ell\nu)+\text{jets}$ production, where in the latter process the charged lepton $\ell$ is not detected or mis-identified.
These backgrounds are evaluated simultaneously using high-statistics control
regions in the 1-lepton and 2-leptons channels.
A dedicated theoretical calculation at next-to-leading order in the phase space relevant for this analysis~\cite{EWVBF} allows the estimation of the total irreducible background with a precision of few-percent.
The multijet background is directly estimated from data.

The final discrimination is obtained by splitting signal and control region events into 16 bins based on \met, the invariant mass of selected dijet pair, their separation in $\phi$, 
and jet multiplicity to maximise the signal/background separation.
Assuming the SM cross-section for the VBF production mode, an observed (expected) upper limit of 0.145 (0.103) at the 95\% CL is placed on \brinv.

\subsection{\zll search}
 
The search targetting the Higgs boson production in association with a \Zboson boson selects events containing a pair of electrons or muons and significant missing transverse momentum.
The two charged leptons are required to have an invariant mass within a narrow window around the \Zboson boson mass for the events to satisfy the signal selection requirements.

The dominant backgrounds for this signature are \(ZZ\), where one of the \Zboson bosons decays into a neutrino--antineutrino pair, and \(WZ\) production.
Contributions from \ttbar and  \(WW\) production are estimated from data, using events with two identified different-flavour charged leptons (electrons and muons).
 
Beyond the signature selections,  sensitivity for the \hinv model is enhanced using a boosted decision tree (BDT) discriminator to improve the separation between signal and background.
A profile likelihood fit to the BDT output distribution results in an observed (expected) upper limit
on \brinv of 0.185 (0.185) at the 95\% CL, assuming the SM production cross-section for this process.

\subsection{\ttH search}
 
The production mode of the Higgs boson in association with a top-quark pair is targetted by reinterpreting the combination of several searches for new phenomena in association with heavy flavour quarks~\cite{SUSY-2018-12,SUSY-2018-07,SUSY-2018-08}. The final states arising from this production mode are characterised by the presence of \(b\)-tagged jets and different charged lepton multiplicities, depending on the decay mode of the two $W$ bosons from the \ttbar decays. In addition, a relevant amount of \met is present, coming from the invisible decay products of the Higgs boson and from neutrinos.
 
A targetted event selection is developed for each lepton multiplicity, resulting in different dominant background contributions from SM processes: \ttbar and $Z(\rightarrow\nu\nu)+\text{jets}$ in the 0-lepton channel, \ttbar in the 1-lepton channel and \(ttZ\) in the 2-lepton channel.
For all the combined analyses, background-enriched selections are defined in order to allow the data to aid in estimating the dominant backgrounds, and validation regions are
used to verify the robustness of these estimates.
 
The combination of the three analyses of each lepton multiplicity, considered in this document as a single combined analysis, places an observed (expected) upper limit on \brinv of 0.376 (0.295) at the 95\% CL, assuming the SM production cross-section for this process.
 
\subsection{\vbfGam search}
 
The VBF topology is further investigated by a dedicated analysis targetting the final states with an emitted photon.
The event signature is characterised by significant missing transverse momentum and one photon in the final state, in addition to a pair of forward jets.
In the SM this topology can arise from \(V\gamma\)+jets production, where \(V\) is either a \Zboson boson decaying into a neutrino pair or a \Wboson boson decaying leptonically, where the charged lepton is missed.
\\
A dense neural network was designed and trained to separate such backgrounds from the \hinv signal by using kinematic properties of the events.
The residual SM contribution to the signal regions is estimated with the aid of specific control regions requiring the presence of electron or muon candidates, to set the normalisation of the MC simulation for \(V\gamma\)+jets processes.
Assuming the SM production cross-section on the signal model, an observed (expected) upper limit on \brinv of 0.375 (0.346) at the 95\% CL is evaluated.

\subsection{\Mj search}
 
The gluon--gluon fusion production mode of the Higgs boson is targetted by a search for new phenomena in events with at least one jet and large missing transverse momentum.
Data are collected with a trigger selection based on the presence of \met and events are vetoed if any charged lepton or photon is reconstructed.
\\
The dominant SM background for this search arises from the irreducible process \(\Zboson\rightarrow\nu\nu\) or \(\Wboson\rightarrow\ell\nu\) in association with jets, where the \Wboson boson decays into either hadronically decaying $\tau$-leptons or undetected electrons or muons. Additional contributions include \ttbar pair or single-top production, diboson production, and non-collision and multijet backgrounds.
The estimate of the major SM processes in the analysis selection is based on a profile likelihood fit to the distribution of the \pT of the system recoiling against the jets reconstructed in the event, performed simultaneously in the signal region and in orthogonal control regions enriched with the targetted backgrounds. Assuming the SM cross-section for Higgs boson gluon--gluon fusion production, an observed (expected) upper limit on \brinv of 0.329 (0.383) at the 95\% CL is achieved.

\subsection{Run~1 combination}
The Run~1 ATLAS \hinv combination utilises 4.7~\ifb~of $pp$ collision data at $\sqrt s=7$ \TeV~and 20.3~\ifb~at $\sqrt s =8$ \TeV~\cite{HIGG-2015-03}. This combination considers inputs from direct detection of \hinv through Higgs bosons produced via VBF or in association with a vector boson $V$, where the vector boson decays either leptonically ($Z\rightarrow\ell\ell$) or hadronically ($W/Z\rightarrow jj$).
All of the signal and control regions are utilized in a maximum-likelihood fit resulting in an observed (expected) upper limit of \brinv$<0.252~(0.265)$ at the 95\% CL. The sensitivity is driven by the VBF channel.


\section{Statistical Model}
 
The statistical combination of the analyses is performed by constructing the product of their respective likelihoods and maximising the resulting profile likelihood ratio~\cite{Cowan:2010js}:
 
\[
\Lambda(\beta;\theta) = \frac{L\left(\beta,\hat{\hat{\theta}}(\beta)\right)}{L\left(\hat{\beta},\hat{\theta}\right)}
\]
 
where \(\beta\) and \(\theta\) are the parameter of interest and the nuisance parameters respectively. In the numerator, the nuisance parameters are set to their fitted values \(\hat{\hat{\theta}}(\beta)\), which maximise the likelihood function for fixed values of the parameter of interest, \(\beta\). In the denominator, both the parameter of interest and the nuisance parameters are set to the values \(\hat{\beta}\) and \(\hat{\theta}\)  which jointly maximise the likelihood. This is done following the implementation described in Ref.~\cite{RooFit,Moneta:2010pm}, with \brinv as the parameter of interest, \(\beta\). Systematic uncertainties are modelled in the likelihood function as nuisance parameters, $\theta$, constrained by Gaussian or log-normal probability density functions.
Upper limits on \brinv are determined following the $CL_s$ formalism~\cite{Read:2002hq} using the profile likelihood ratio as a test statistic.
 
\subsection{Uncertainty correlation in Run~2 combination}
In the combination of Run 2 results, most experimental systematic uncertainties, as well as the uncertainty
on the integrated luminosity and the modelling of additional $pp$ collisions in the same and neighbouring
bunch crossings (pile-up), are correlated across all search channels.
The assessment of some of the uncertainties associated with the calibration of the jet energy scale (JES) and
the jet energy resolution varies between the different analyses in terms of jet reconstruction algorithms and
parameterisation choices. For this reason, the uncertainty components  stemming from  identical methodologies are presumed to be correlated,
while the rest of the uncertainties are treated as uncorrelated.
Finally, a few experimental systematic uncertainties that are tightly constrained in a given analysis are not correlated
in order not to introduce any potential phase space specific biases. 
The impact of these assumptions on the combined result is estimated by using alternative correlation models
and found to have an absolute effect on the \brinv limit of the order of 0.003.
 
The uncertainties related to background predictions are considered to be uncorrelated among analyses due to the different
nature of the leading backgrounds, the variety of kinematic phase space covered by the various analyses, and the usage of data-driven techniques.
The systematic uncertainties in the prediction of Higgs boson production follow the recommendations in Ref.~\cite{deFlorian:2016spz}.
Variations connected to the choice of parton distribution functions (PDF) are considered as correlated among channels while effects of missing higher-order contributions (estimates through variations of factorisation and renormalisation scales) and parton shower/hadronisation models are considered independently for each Higgs boson production mode and therefore uncorrelated across the analyses.

\subsection{Uncertainty correlation in Run~1 and Run~2 combination}
 
The Run 2 result described above is combined with the Run~1 searches for \hinv decays.
The adopted correlation scheme follows closely the statistical combination of the partial Run~2 results with the Run~1 combination~\cite{HIGG-2018-54}.
 
The correlation schemes of the individual Run~1 and Run~2 combinations are preserved when combined together.
Due to the differences between the detector layouts and data-taking conditions, reconstruction algorithms, which are calibrated using data,
and treatment of systematic uncertainties, the correlations between the two LHC runs are not clearly identifiable. Hence, no correlations between Run~1 and~2 are assumed for most instrumental uncertainties. Exceptions are made for uncertainties related to the modelling of the calorimeter response dependence on jet flavour and pile-up, the calibration of the JES across different $\eta$ regions, and the uncertainties related to the JES of $b$-quark jets. Such components are treated as correlated given that the same methodology was applied to compute them in both of the datasets.
 
Background modelling uncertainties are considered to be uncorrelated in order to reflect improvements in the MC simulation tools and general theory predictions that have evolved significantly since Run~1, both on the side of the hard process simulation and on the side of the parton shower and hadronisation models.
For similar reasons, the signal modelling uncertainties are considered uncorrelated between the Run~1 and Run~2 combinations.
 
The result of the combination shows little sensitivity to the exact correlation scheme between the Run~1 and Run~2 results due to the dominant weight of the
latter.


\section{Results}

The value of twice the negative logarithmic profile likelihood ratio $-2\,\ln(\Lambda)(\brinv;\theta)$ as a function of \brinv of the individual analyses and of the combined Run 2 result are shown in Figure~\ref{fig:lhscan_rtwo} (left). The combined best-fit value for \brinv is $0.04 \pm 0.04$. 
Good agreement among the best fit values of the individual analyses, reported in Table~\ref{tab:resultR2}, is observed.
 
\begin{figure}[ht!b]
\begin{center}
\includegraphics[width=0.48\textwidth]{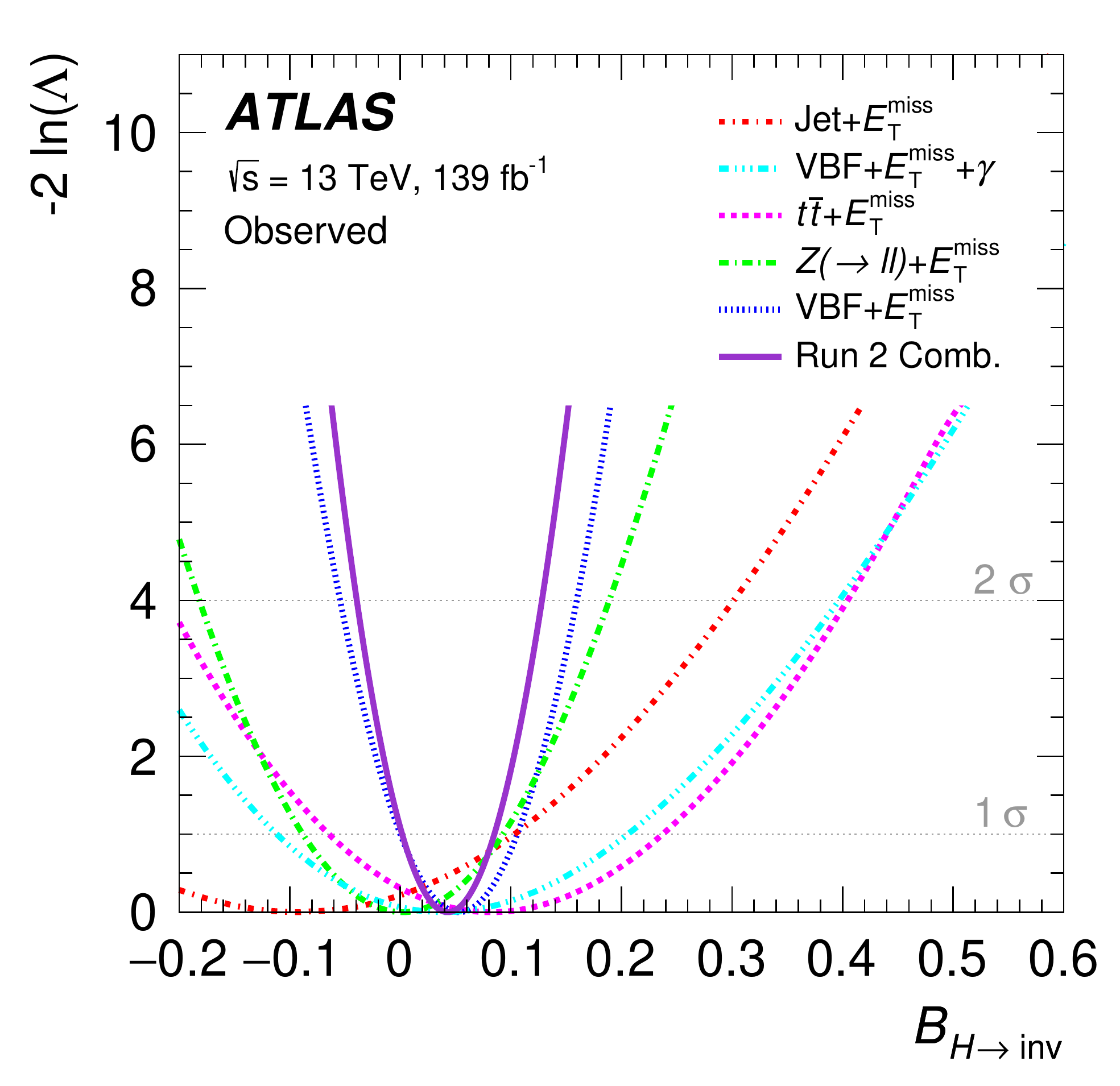}
\includegraphics[width=0.48\textwidth]{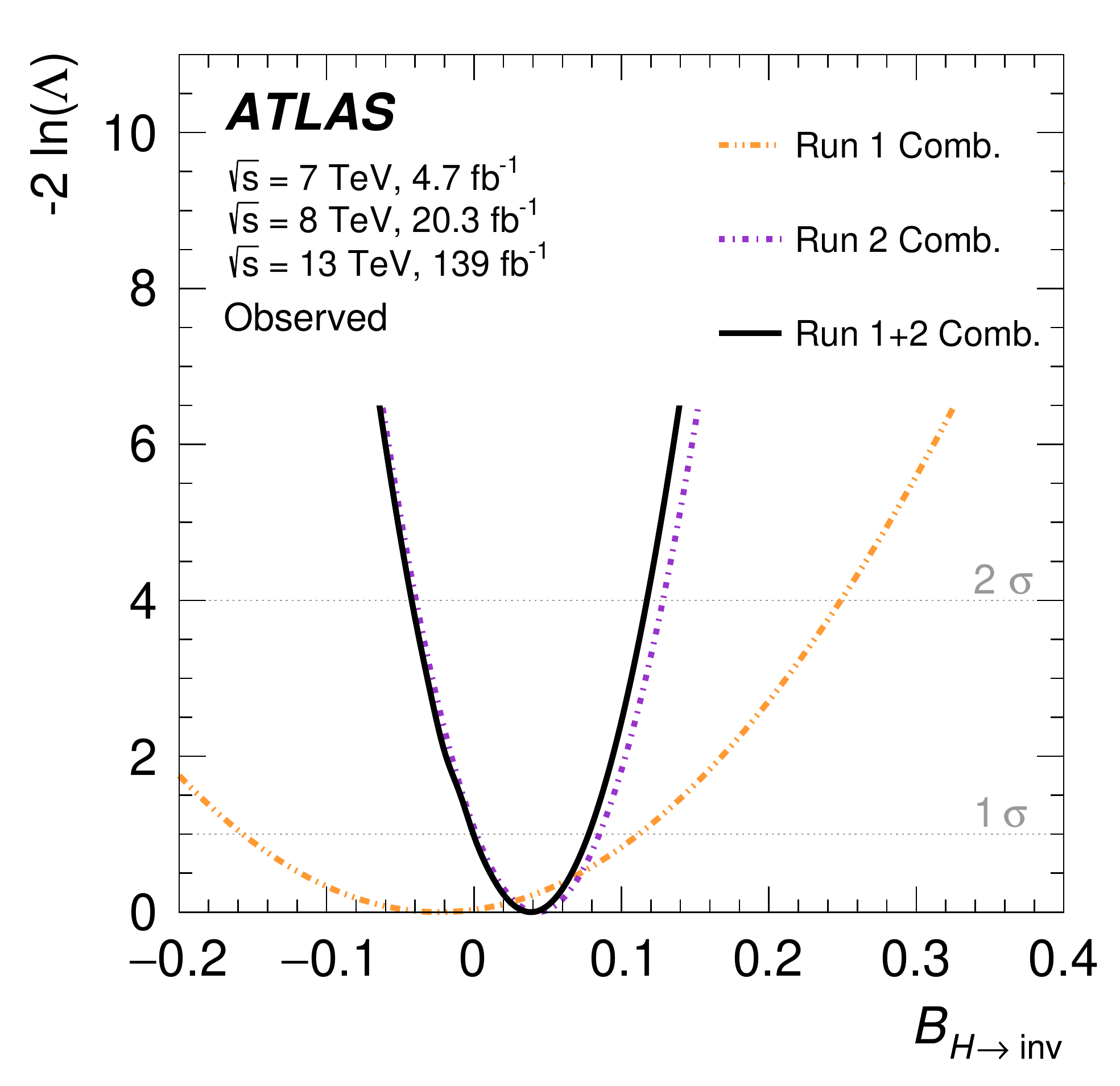}
\end{center}
\caption{
The observed value of $-2\,\ln(\Lambda)$ as a function of \brinv for the individual Run~2
analyses and their combination (left) and the Run~2 combination together with the Run~1 combination and the total Run~1+2 combination (right).
\label{fig:lhscan_rtwo}
}
\end{figure}

The best-fit values for \brinv together with the 95\% CL expected and observed upper limits for each individual Run~2 analysis and their combination are also shown in Table~\ref{tab:resultR2}.
An upper limit of 0.113 is observed for the combined Run~2 data, while an upper limit of 0.080 was expected in the case of no observed excess in data.
Relative to the most sensitive single analysis, the VBF final state, the Run 2 combination brings a relative sensitivity improvement of 22\%.
 
Overall, the leading systematic uncertainty  of the result  is due to the modelling uncertainties of the $W$/$Z$+jets prediction in the \vbf search~\cite{EXOT-2020-11}.
Subdominant uncertainties with similar contribution are related to the statistical precision of the data sample; the number of simulated MC events, in particular for the $W$/$Z$+jets process; the reconstruction and identification of jets and leptons; and the modelling of background processes other than from $W/Z$+jets production.

\begin{table} [hbp!t]
\begin{center}
\caption{
Best fit value, observed and expected 95\% upper limit on \brinv for each individual Run~2 analysis, their combination, the Run~1 combination and the full Run~1+2 combination.}
\label{tab:resultR2}
\renewcommand{\arraystretch}{1.4}
\begin{tabular}{l | c || c | c }
\hline
\hline
Analysis  & Best fit \brinv & Observed 95\% U.L. & Expected 95\% U.L.  \\
\hline
\Mj 			&  $          -0.09 ^{+0.19}_{-0.20}$  & 0.329 & $0.383^{+0.157}_{-0.107}$ \\
\vbfGam		        &  $\phantom{-}0.04 ^{+0.17}_{-0.15}$  & 0.375 & $0.346^{+0.151}_{-0.097}$ \\
\ttH 			&  $\phantom{-}0.08 \pm 0.15$       & 0.376 & $0.295^{+0.125}_{-0.083}$ \\
\zll 			&  $\phantom{-}0.00 \pm 0.09$       & 0.185 & $0.185^{+0.078}_{-0.052}$ \\
\vbf 			&  $\phantom{-}0.05 \pm 0.05$       & 0.145 & $0.103^{+0.041}_{-0.028}$ \\
\hline
Run 2 Comb.             &  $\phantom{-}0.04 \pm 0.04$      & 0.113 & $0.080^{+0.031}_{-0.022}$ \\
\hline
Run 1 Comb.             &  $          -0.02 ^{+0.14}_{-0.13}$ & 0.252 & $0.265^{+0.105}_{-0.074}$ \\
\hline
Run 1+2 Comb.           &  $\phantom{-}0.04 \pm 0.04$      & 0.107 & $0.077^{+0.030}_{-0.022}$ \\
\hline
\hline
\end{tabular}
\renewcommand{\arraystretch}{1.0}
\end{center}
\end{table}
 
The observed $-2\,\ln(\Lambda)(\brinv;\theta)$ scan of the combined Run~1+2 result is represented in Figure~\ref{fig:lhscan_rtwo} (right), alongside the individual Run~1 and Run~2 combinations.
A best-fit value of ${\brinv=0.04 \pm 0.04}$ is obtained for the Run~1+2 combination, corresponding to an observed~(expected) upper limit of \brinv $<~0.107~(0.077)$ at the 95\%~CL.
The result is dominated by the Run~2 analysis with the addition of Run~1 combination improving the expected relative sensitivity by 4\%.
 
The overall picture of the most relevant sources of uncertainty in the Run~1+Run~2 combination is very similar to that of the Run~2 combination. The upper limit would improve by 50\% if all sources of systematic uncertainties were ignored.
 
The upper limits for each individual Run~2 analysis, their combination, the Run~1 combination and the overall Run~1+2 combined result are summarised in Figure~\ref{fig:R2UL}.
The current combination improves the constraints on \brinv by more than a factor of two as compared to the previous ATLAS combination from Run~1 and partial Run~2 results~\cite{HIGG-2018-54}.
 
\begin{figure}[h!tbp]
\centering
\includegraphics[scale=0.65]{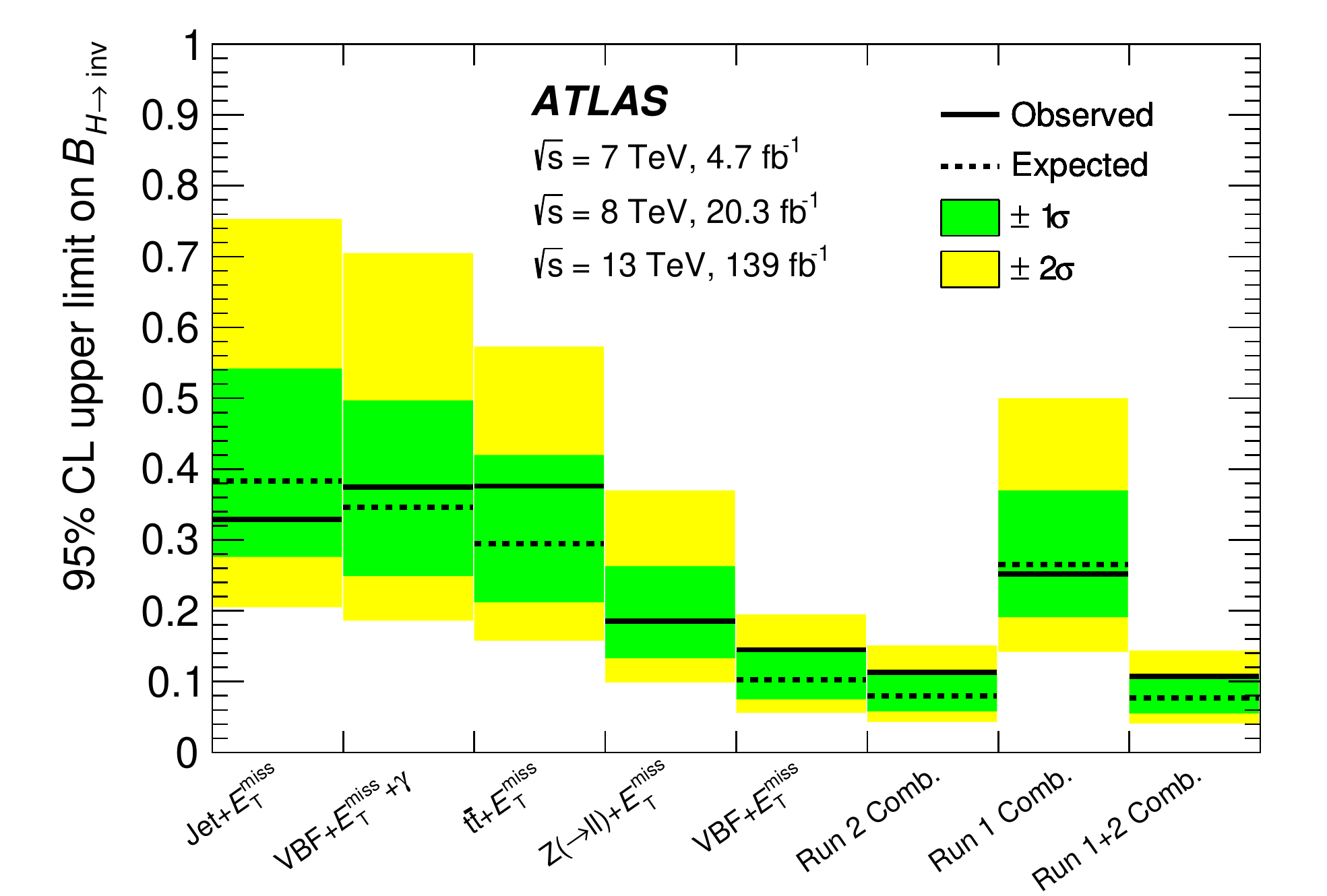}
\caption{
The observed and expected  upper limits on \brinv at 95\% CL for the Run~2 analyses targetting the \mj, \vbfGam, \ttH, \zll, \vbf final states and their combination, the Run~1 combination and the full Run~1+2 result; the \(1\sigma\) and \(2\sigma\) contours of the expected limit distribution are also shown. 
}
\label{fig:R2UL}
\end{figure}


\section{Comparison to direct dark matter detection experiments}
 
The combined observed Run~1+2 upper limit on \brinv can be converted into a limit on the
spin-independent scattering cross-section of a weakly interacting massive particle~(WIMP) and a nucleon~\cite{Eboli:2000ze, Kanemura:2010sh,Fox_2012,De_Simone_2014} (\sigmawn), to allow the comparison of the results with the ones from experiments based on different detector technologies.
The translation is performed in the context of Higgs portal models~\cite{Patt:2006fw,Djouadi:2012zc} using an effective field theory framework,
where the mediator of new interactions is assumed to be above the TeV-level and therefore well above the scale probed at the Higgs boson mass.
The approach assumes that  Higgs boson decays into a pair of WIMP particles are kinematically possible ($m_{\text{WIMP}} < m_{H}/2$)
and that the WIMP particle is either a scalar, a Majorana fermion, or a vector-like
state\footnote{The value of $f_N=0.308\pm0.018$~\cite{Hoferichter:2017olk} is used as nuclear form factor.}.
In addition, in the case of vectorial DM states, various ultraviolet-complete (UV) models were proposed~\cite{Baek:2014jga,Arcadi:2020jqf,Zaazoua:2021xls}.
In such scenarios, the vector DM candidate is introduced as a gauge field of a U(1)$^\prime$ group which extends
the SM symmetry group and a dark Higgs sector is added to generate the vector boson mass via the Higgs spontaneous symmetry breaking mechanism.
This adds at least two free parameters to the model: the mass $m_{2}$ of the additional dark Higgs boson and its mixing angle $\alpha$ with the SM Higgs boson.
 
\begin{figure}[ht!b]
\centering
\includegraphics[width=0.7\textwidth]{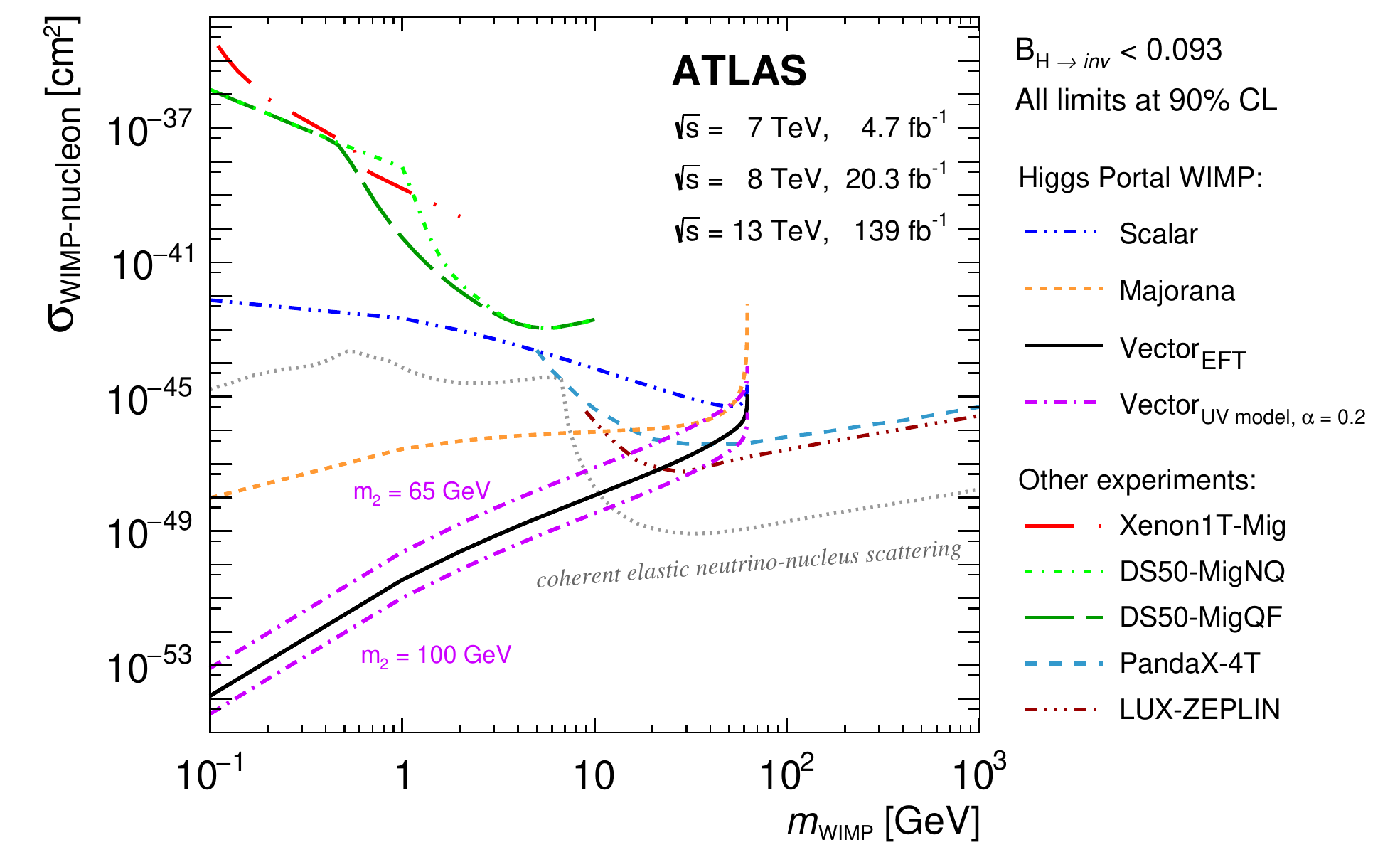}
\caption{
Upper limit at the 90\% CL on the spin-independent WIMP-nucleon scattering cross-section as a function of the WIMP mass for direct detection experiments and the interpretation of the \hinv combination result in the context of Higgs portal models considering scalar, Majorana and vector WIMP hypotheses. For the vector case, results from UV-complete models are shown (pink curves) for two representative values for the mass of the predicted Dark Higgs particle ($m_{2}$) and a mixing angle $\alpha$=0.2. The uncertainties from the nuclear form factor are smaller than the line thickness. 
Direct detection results are taken from Refs.~\cite{Agnes_2018, Meng_2021, LZ:2022ufs,XENON:2019zpr}.
The neutrino floor for coherent elastic neutrino-nucleus scattering (dotted gray line) is taken from Refs.~\cite{Billard_2014, Ruppin_2014}, which assume that germanium is the target over the whole WIMP mass range.
The regions above the limit contours are excluded in the range shown in the plot.
}
\label{fig:dd}
\end{figure}

The constraint from the combined observed Run 1+2 exclusion limit of $\brinv < 0.093$ at 90\% CL is compared to the results from representative direct DM detection experiments~\cite{Agnes_2018, Meng_2021, LZ:2022ufs,XENON:2019zpr} in Figure~\ref{fig:dd}.
The excluded \sigmawn values range from $10^{-45}~$cm$^{2}$ to $10^{-42}~$cm$^{2}$ in the scalar WIMP scenario.
In the Majorana fermion WIMP case, the effective coupling is reduced by a factor $m_H^2$~\cite{HIGG-2013-16}, excluding cross-section values down to $2 \times 10^{-47}~$cm$^{2}$ for low WIMP masses; \sigmawn values down to $10^{-54}~$cm$^{2}$ can be excluded for the vector WIMP hypothesis.
For UV-complete models, Figure~\ref{fig:dd} also shows the upper limit cross-section behaviour for a mixing angle $\alpha=0.2$ and for masses of the dark Higgs particle equal to 65~GeV and 100~GeV corresponding to the worst and best limit for a scan of $m_{2}$ in the range [65, 1000]~\GeV~\cite{Zaazoua:2021xls}.
This comparison illustrates the complementarity in coverage by the direct-detection experiments and the searches at colliders, such as the presented analysis.


\FloatBarrier
 
\section{Conclusion}
In summary, searches for invisible decays of the Higgs boson using 139~\ifb\ of $pp$ collision data at $\sqrt{s} = 13~\TeV$ recorded in Run 2 of the LHC in several Higgs boson production topologies were statistically combined assuming SM Higgs boson production. An upper limit on the invisible Higgs boson branching ratio of
$
\brinv{}<\result
$
is observed~(expected) at the 95\% CL. A statistical combination of this result with the combination of \hinv searches using up to 4.7~\ifb\ of $pp$ collision data at $\sqrt{s} = 7$~\TeV\ and up to 20.3~\ifb\ at 8~TeV collected in Run~1 of the LHC yields an observed~(expected) upper limit of
$
\brinv{}<\resultall
$
at the 95\% CL.
The combined Run~1+2 result is translated into upper limits on the WIMP-nucleon scattering cross-section for Higgs portal models.
The derived limits on \sigmawn range down to $10^{-45}~$cm$^{2}$ (scalar), $2\times 10^{-47}~$cm$^{2}$ (Majorana) and $10^{-54}~$cm$^{2}$ (vector), highlighting the complementarity of DM searches at the LHC and direct detection experiments.
 
\section*{Acknowledgements}


We thank CERN for the very successful operation of the LHC, as well as the
support staff from our institutions without whom ATLAS could not be
operated efficiently.
 
We acknowledge the support of
ANPCyT, Argentina;
YerPhI, Armenia;
ARC, Australia;
BMWFW and FWF, Austria;
ANAS, Azerbaijan;
CNPq and FAPESP, Brazil;
NSERC, NRC and CFI, Canada;
CERN;
ANID, Chile;
CAS, MOST and NSFC, China;
Minciencias, Colombia;
MEYS CR, Czech Republic;
DNRF and DNSRC, Denmark;
IN2P3-CNRS and CEA-DRF/IRFU, France;
SRNSFG, Georgia;
BMBF, HGF and MPG, Germany;
GSRI, Greece;
RGC and Hong Kong SAR, China;
ISF and Benoziyo Center, Israel;
INFN, Italy;
MEXT and JSPS, Japan;
CNRST, Morocco;
NWO, Netherlands;
RCN, Norway;
MEiN, Poland;
FCT, Portugal;
MNE/IFA, Romania;
MESTD, Serbia;
MSSR, Slovakia;
ARRS and MIZ\v{S}, Slovenia;
DSI/NRF, South Africa;
MICINN, Spain;
SRC and Wallenberg Foundation, Sweden;
SERI, SNSF and Cantons of Bern and Geneva, Switzerland;
MOST, Taiwan;
TENMAK, T\"urkiye;
STFC, United Kingdom;
DOE and NSF, United States of America.
In addition, individual groups and members have received support from
BCKDF, CANARIE, Compute Canada and CRC, Canada;
PRIMUS 21/SCI/017 and UNCE SCI/013, Czech Republic;
COST, ERC, ERDF, Horizon 2020 and Marie Sk{\l}odowska-Curie Actions, European Union;
Investissements d'Avenir Labex, Investissements d'Avenir Idex and ANR, France;
DFG and AvH Foundation, Germany;
Herakleitos, Thales and Aristeia programmes co-financed by EU-ESF and the Greek NSRF, Greece;
BSF-NSF and MINERVA, Israel;
Norwegian Financial Mechanism 2014-2021, Norway;
NCN and NAWA, Poland;
La Caixa Banking Foundation, CERCA Programme Generalitat de Catalunya and PROMETEO and GenT Programmes Generalitat Valenciana, Spain;
G\"{o}ran Gustafssons Stiftelse, Sweden;
The Royal Society and Leverhulme Trust, United Kingdom.
 
The crucial computing support from all WLCG partners is acknowledged gratefully, in particular from CERN, the ATLAS Tier-1 facilities at TRIUMF (Canada), NDGF (Denmark, Norway, Sweden), CC-IN2P3 (France), KIT/GridKA (Germany), INFN-CNAF (Italy), NL-T1 (Netherlands), PIC (Spain), ASGC (Taiwan), RAL (UK) and BNL (USA), the Tier-2 facilities worldwide and large non-WLCG resource providers. Major contributors of computing resources are listed in Ref.~\cite{ATL-SOFT-PUB-2021-003}.


\appendix
 
\printbibliography

@article{Ruppin_2014,
	doi = {10.1103/physrevd.90.083510}, 
	%url = {https://arxiv.org/abs/1408.3581},
        eprint         = "1408.3581",
        archivePrefix  = "arXiv",
        XprimaryClass   = "hep-ph",
	year = 2014,
	month = {oct},
	publisher = {American Physical Society ({APS})},
	volume = {90},
	number = {8},
	author = {F. Ruppin and J. Billard and E. Figueroa-Feliciano and L. Strigari},
	title = {Complementarity of dark matter detectors in light of the neutrino background},
	journal = {Physical Review D}
}

@Article{Moneta:2010pm,
     author    = "Moneta, Lorenzo and others",
     title     = "{The RooStats project}",
     journal   = "PoS",
     volume    = "ACAT2010",
     year      = "2010",
     pages     = "057",
     eprint    = "1009.1003",
     archivePrefix = "arXiv",
     primaryClass  =  "physics.data-an",
     SLACcitation  = "%%CITATION = 1009.1003;%%"
}

@misc{Patt:2006fw,
      author         = "Patt, Brian and Wilczek, Frank",
      title          = "{Higgs-field portal into hidden sectors}",
      year           = "2006",
      eprint         = "hep-ph/0605188",
      archivePrefix  = "arXiv",
      primaryClass   = "hep-ph",
      reportNumber   = "MIT-CTP-3745",
      SLACcitation   = "%%CITATION = HEP-PH/0605188;%%"
}

@article{Billard_2014,
	doi = {10.1103/physrevd.89.023524},  
	%url = {https://arxiv.org/abs/1307.5458},
        eprint         = "1307.5458",
        archivePrefix  = "arXiv",
        primaryClass   = "hep-ph",
	year = 2014,
	month = {jan},
	publisher = {American Physical Society ({APS})},
	volume = {89},
	number = {2},
	author = {J. Billard and E. Figueroa-Feliciano and L. Strigari},
	title = {Implication of neutrino backgrounds on the reach of next generation dark matter direct detection experiments},
	journal = {Physical Review D}
}

@article{Zaazoua:2021xls,
    author = "Zaazoua, Mohamed and Truong, Loan and Assamagan, K\'et\'evi A. and Fassi, Farida",
    title = "{Higgs Portal Vector Dark Matter Interpretation: Review of Effective Field Theory Approach and Ultraviolet Complete Models}",
    eprint = "2107.01252",
    archivePrefix = "arXiv",
    primaryClass = "hep-ph",
    doi = "10.31526/lhep.2022.270",
    month = "7",
    year = "2022",
    journal = "Letters in High Energy Physics",
    %volume  = "579",
    pages   = "270"
}

@article{Agnes_2018,
%	doi = {10.1103/physrevlett.121.081307}, 
	%url = {https://arxiv.org/abs/1802.06994},
%        eprint         = "1802.06994",
%        archivePrefix  = "arXiv",
%        XprimaryClass   = "astro-ph.HE",
%	year = 2018,
%	month = {aug},
%	publisher = {American Physical Society ({APS})},
%	volume = {121},
%	number = {8},
%	author = {P. Agnes and I.{\hspace{0.167em}
%}F.{\hspace{0.167em}}M. Albuquerque and T. Alexander and A.{\hspace{0.167em}}K. Alton and G.{\hspace{0.167em}}R. Araujo and D.{\hspace{0.167em}}M. Asner and M. Ave and H.{\hspace{0.167em}}O. Back and B. Baldin and G. Batignani and K. Biery and V. Bocci and G. Bonfini and W. Bonivento and B. Bottino and F. Budano and S. Bussino and M. Cadeddu and M. Cadoni and F. Calaprice and A. Caminata and N. Canci and A. Candela and M. Caravati and M. Cariello and M. Carlini and M. Carpinelli and S. Catalanotti and V. Cataudella and P. Cavalcante and S. Cavuoti and R. Cereseto and A. Chepurnov and C. Cical{\`{o}} and L. Cifarelli and A.{\hspace{0.167em}}G. Cocco and G. Covone and D. D'Angelo and M. D'Incecco and D. D'Urso and S. Davini and A. De Candia and S. De Cecco and M. De Deo and G. De Filippis and G. De Rosa and M. De Vincenzi and P. Demontis and A.{\hspace{0.167em}}V. Derbin and A. Devoto and F. Di Eusanio and G. Di Pietro and C. Dionisi and M. Downing and E. Edkins and A. Empl and A. Fan and G. Fiorillo and K. Fomenko and D. Franco and F. Gabriele and A. Gabrieli and C. Galbiati and P. Garcia Abia and Chiara Ghiano and S. Giagu and C. Giganti and G.{\hspace{0.167em}}K. Giovanetti and O. Gorchakov and A.{\hspace{0.167em}}M. Goretti and F. Granato and M. Gromov and M. Guan and Y. Guardincerri and M. Gulino and B.{\hspace{0.167em}}R. Hackett and M.{\hspace{0.167em}}H. Hassanshahi and K. Herner and B. Hosseini and D. Hughes and P. Humble and E.{\hspace{0.167em}}V. Hungerford and Al. Ianni and An. Ianni and V. Ippolito and I. James and T.{\hspace{0.167em}}N. Johnson and Y. Kahn and K. Keeter and C.{\hspace{0.167em}}L. Kendziora and I. Kochanek and G. Koh and D. Korablev and G. Korga and A. Kubankin and M. Kuss and M. La Commara and M. Lai and X. Li and M. Lisanti and M. Lissia and B. Loer and G. Longo and Y. Ma and A.{\hspace{0.167em}}A. Machado and I.{\hspace{0.167em}}N. Machulin and A. Mandarano and L. Mapelli and S.{\hspace{0.167em}}M. Mari and J. Maricic and C.{\hspace{0.167em}}J. Martoff and A. Messina and P.{\hspace{0.167em}}D. Meyers and R. Milincic and S. Mishra-Sharma and A. Monte and M. Morrocchi and B.{\hspace{0.167em}}J. Mount and V.{\hspace{0.167em}}N. Muratova and P. Musico and R. Nania and A. Navrer Agasson and A.{\hspace{0.167em}}O. Nozdrina and A. Oleinik and M. Orsini and F. Ortica and L. Pagani and M. Pallavicini and L. Pandola and E. Pantic and E. Paoloni and F. Pazzona and K. Pelczar and N. Pelliccia and V. Pesudo and A. Pocar and S. Pordes and S.{\hspace{0.167em}}S. Poudel and D.{\hspace{0.167em}}A. Pugachev and H. Qian and F. Ragusa and M. Razeti and A. Razeto and B. Reinhold and A.{\hspace{0.167em}}L. Renshaw and M. Rescigno and Q. Riffard and A. Romani and B. Rossi and N. Rossi and D. Sablone and O. Samoylov and W. Sands and S. Sanfilippo and M. Sant and R. Santorelli and C. Savarese and E. Scapparone and B. Schlitzer and E. Segreto and D.{\hspace{0.167em}}A. Semenov and A. Shchagin and A. Sheshukov and P.{\hspace{0.167em}}N. Singh and M.{\hspace{0.167em}}D. Skorokhvatov and O. Smirnov and A. Sotnikov and C. Stanford and S. Stracka and G.{\hspace{0.167em}}B. Suffritti and Y. Suvorov and R. Tartaglia and G. Testera and A. Tonazzo and P. Trinchese and E.{\hspace{0.167em}}V. Unzhakov and M. Verducci and A. Vishneva and B. Vogelaar and M. Wada and T.{\hspace{0.167em}}J. Waldrop and H. Wang and Y. Wang and A.{\hspace{0.167em}}W. Watson and S. Westerdale and M.{\hspace{0.167em}}M. Wojcik and M. Wojcik and X. Xiang and X. Xiao and C. Yang and Z. Ye and C. Zhu and A. Zichichi and G. Zuzel and},
%	title = {Low-Mass Dark Matter Search with the {DarkSide}-50 Experiment},
%	journal = {Physical Review Letters}
%}

@article{Meng_2021,
	doi = {10.1103/physrevlett.127.261802},  
	%url = {https://arxiv.org/abs/2107.13438},
        eprint         = "2107.13438",
        archivePrefix  = "arXiv",
        XprimaryClass  = "hep-ex",
	year = 2021,
	month = {dec},
	publisher = {American Physical Society ({APS})},
	volume = {127},
	number = {26},
	author = {Yue Meng and Zhou Wang and Yi Tao and Abdusalam Abdukerim and Zihao Bo and Wei Chen and Xun Chen and Yunhua Chen and Chen Cheng and Yunshan Cheng and Xiangyi Cui and Yingjie Fan and Deqing Fang and Changbo Fu and Mengting Fu and Lisheng Geng and Karl Giboni and Linhui Gu and Xuyuan Guo and Ke Han and Changda He and Jinrong He and Di Huang and Yanlin Huang and Zhou Huang and Ruquan Hou and Xiangdong Ji and Yonglin Ju and Chenxiang Li and Mingchuan Li and Shu Li and Shuaijie Li and Qing Lin and Jianglai Liu and Xiaoying Lu and Lingyin Luo and Wenbo Ma and Yugang Ma and Yajun Mao and Nasir Shaheed and Xuyang Ning and Ningchun Qi and Zhicheng Qian and Xiangxiang Ren and Changsong Shang and Guofang Shen and Lin Si and Wenliang Sun and Andi Tan and Anqing Wang and Meng Wang and Qiuhong Wang and Shaobo Wang and Siguang Wang and Wei Wang and Xiuli Wang and Mengmeng Wu and Weihao Wu and Jingkai Xia and Mengjiao Xiao and Xiang Xiao and Pengwei Xie and Binbin Yan and Xiyu Yan and Jijun Yang and Yong Yang and Chunxu Yu and Jumin Yuan and Ying Yuan and Dan Zhang and Minzhen Zhang and Peng Zhang and Tao Zhang and Li Zhao and Qibin Zheng and Jifang Zhou and Ning Zhou and Xiaopeng Zhou and Yong Zhou and},
	title = {Dark Matter Search Results from the {PandaX}-4T Commissioning Run},
	journal = {Physical Review Letters}
}

@article{LZ:2022ufs,
    author = "Aalbers, J. and others",
    collaboration = "LZ",
    title = "{First Dark Matter Search Results from the LUX-ZEPLIN (LZ) Experiment}",
    eprint = "2207.03764",
    archivePrefix = "arXiv",
    primaryClass = "hep-ex",
    month = "7",
    year = "2022"
}

@article{Fox_2012,
   title={Missing energy signatures of dark matter at the LHC},
   volume={85},
   ISSN={1550-2368},
   %url={http://dx.doi.org/10.1103/PhysRevD.85.056011},
   DOI={10.1103/physrevd.85.056011},
   eprint         = "1109.4398",
   archivePrefix  = "arXiv",
   XprimaryClass  = "hep-ph",
   number={5},
   journal={Physical Review D},
   publisher={American Physical Society (APS)},
   author={Fox, Patrick J. and Harnik, Roni and Kopp, Joachim and Tsai, Yuhsin},
   year={2012},
   month={Mar}
}

@article{De_Simone_2014,
   title={Benchmarks for dark matter searches at the LHC},
   volume={2014},
   %ISSN={1029-8479},
   %url={http://dx.doi.org/10.1007/JHEP06(2014)081},
   DOI={10.1007/jhep06(2014)081},
   %number={6},
   journal={JHEP},
   volume         = "06",
   year           = "2014",
   pages          = "081",
   publisher={Springer Science and Business Media LLC},
   eprint         = "1402.6287",
   archivePrefix  = "arXiv",
   XprimaryClass   = "hep-ph",
   author={De Simone, Andrea and Giudice, Gian Francesco and Strumia, Alessandro},
   year={2014},
   month={Jun}
}

@article{Kanemura:2010sh,
      author         = "Kanemura, Shinya and Matsumoto, Shigeki and Nabeshima,
                        Takehiro and Okada, Nobuchika",
      title          = "{Can WIMP dark matter overcome the nightmare scenario?}",
      journal        = "Phys. Rev. D",
      volume         = "82",
      year           = "2010",
      pages          = "055026",
      doi            = "10.1103/PhysRevD.82.055026",
      eprint         = "1005.5651",
      archivePrefix  = "arXiv",
      primaryClass   = "hep-ph",
      reportNumber   = "UT-HET-039",
      SLACcitation   = "%%CITATION = ARXIV:1005.5651;%%"
}

@article{Eboli:2000ze,
      author         = "Eboli, Oscar J. P. and Zeppenfeld, D.",
      title          = "{Observing an invisible Higgs boson}",
      journal        = "Phys. Lett. B",
      volume         = "495",
      year           = "2000",
      pages          = "147-154",
      doi            = "10.1016/S0370-2693(00)01213-2",
      eprint         = "hep-ph/0009158",
      archivePrefix  = "arXiv",
      primaryClass   = "hep-ph",
      reportNumber   = "MADPH-00-1191, IFT-P-076-2000",
      SLACcitation   = "%%CITATION = HEP-PH/0009158;%%"
}

@article{Djouadi:2011aa,
      author         = "Djouadi, Abdelhak and Lebedev, Oleg and Mambrini, Yann
                        and Quevillon, Jeremie",
      title          = "{Implications of LHC searches for Higgs--portal dark
                        matter}",
      journal        = "Phys. Lett. B",
      volume         = "709",
      year           = "2012",
      pages          = "65-69",
      doi            = "10.1016/j.physletb.2012.01.062",
      eprint         = "1112.3299",
      archivePrefix  = "arXiv",
      primaryClass   = "hep-ph",
      reportNumber   = "DESY-11-240, LPT-ORSAY-11-117",
      SLACcitation   = "%%CITATION = ARXIV:1112.3299;%%"
}

@article{Djouadi:2012zc,
      author         = "Djouadi, Abdelhak and Falkowski, Adam and Mambrini, Yann
                        and Quevillon, Jeremie",
      title          = "{Direct detection of Higgs-portal dark matter at the
                        LHC}",
      journal        = "Eur. Phys. J. C",
      volume         = "73",
      year           = "2013",
      number         = "6",
      pages          = "2455",
      doi            = "10.1140/epjc/s10052-013-2455-1",
      eprint         = "1205.3169",
      archivePrefix  = "arXiv",
      primaryClass   = "hep-ph",
      reportNumber   = "LPT-ORSAY-12-42",
      SLACcitation   = "%%CITATION = ARXIV:1205.3169;%%"
}

@article{Arcadi:2020jqf,
    author = "Arcadi, Giorgio and Djouadi, Abdelhak and Kado, Marumi",
    title = "{The Higgs-portal for vector Dark Matter and the Effective Field Theory approach: a reappraisal}",
    eprint = "2001.10750",
    archivePrefix = "arXiv",
    primaryClass = "hep-ph",
    doi = "10.1016/j.physletb.2020.135427",
    journal = "Phys. Lett. B",
    volume = "805",
    pages = "135427",
    year = "2020"
}

@article{Baek:2014jga,
    author = "Baek, Seungwon and Ko, P. and Park, Wan-Il",
    title = "{Invisible Higgs decay width versus dark matter direct detection cross section in Higgs portal dark matter models}",
    eprint = "1405.3530",
    archivePrefix = "arXiv",
    primaryClass = "hep-ph",
    doi = "10.1103/PhysRevD.90.055014",
    journal = "Phys. Rev. D",
    volume = "90",
    number = "5",
    pages = "055014",
    year = "2014"
}

@article{Hoferichter:2017olk,
      author         = "Hoferichter, Martin and Klos, Philipp and Men\'e{}ndez,
                        Javier and Schwenk, Achim",
      title          = "{Improved Limits for Higgs-Portal Dark Matter from LHC Searches}",
      journal        = "Phys. Rev. Lett.",
      volume         = "119",
      year           = "2017",
      number         = "18",
      pages          = "181803",
      doi            = "10.1103/PhysRevLett.119.181803",
      eprint         = "1708.02245",
      archivePrefix  = "arXiv",
      primaryClass   = "hep-ph",
      reportNumber   = "INT-PUB-17-031",
      SLACcitation   = "%%CITATION = ARXIV:1708.02245;%%"
}

@misc{EWVBF,
  doi = {10.48550/arXiv.2204.07652},
  %url = {https://arxiv.org/abs/2204.07652},
  author = {Lindert, Jonas M. and Pozzorini, Stefano and Schönherr, Marek},
  keywords = {High Energy Physics - Phenomenology (hep-ph), High Energy Physics - Experiment (hep-ex), FOS: Physical sciences, FOS: Physical sciences},
  title = {Precise predictions for V+2 jet backgrounds in searches for invisible Higgs decays},
  eprint         = "2204.07652",
  archivePrefix  = "arXiv",
  primaryClass   = "hep-ph",
  reportNumber   = "IPPP/22/22, ZU-TH 15/22",
  year = {2022}
  %publisher = {arXiv},
  %year = {2022},
  %copyright = {arXiv.org perpetual, non-exclusive license}
}

@article{Zwicky:1933gu,
       author         = "Zwicky, F.",
       title          = "{Die Rotverschiebung von extragalaktischen Nebeln}",
       journal        = "Helv.~Phys.~Acta",
       volume         = "6",
       pages          = "110-127",
       year           = "1933",
       SLACcitation   = "%%CITATION = HPACA,6,110;%%",
}

@article{Planck:2015fie,
    author = "{Planck Collaboration}",
    collaboration = "Planck",
    title = "{Planck 2015 results. XIII. Cosmological parameters}",
    eprint = "1502.01589",
    archivePrefix = "arXiv",
    primaryClass = "astro-ph.CO",
    doi = "10.1051/0004-6361/201525830",
    journal = "Astron. Astrophys.",
    volume = "594",
    pages = "A13",
    year = "2016"
 }

@Article{DMreview,
      author    = "Bertone, Gianfranco and Hooper, Dan and Silk, Joseph",
      title     = "{Particle dark matter: Evidence, candidates and
                   constraints}",
      journal   = "Phys. Rept.",
      volume    = "405",
      year      = "2005",
      pages     = "279-390",
      doi       = "10.1016/j.physrep.2004.08.031",
      eprint    = "hep-ph/0404175",
      archivePrefix = "arXiv",
      SLACcitation  = "%%CITATION = HEP-PH/0404175;%%"

 }

@Article{Komatsu:2010fb,
      author         = "{E. Komatsu et al.}",
      title          = "{SEVEN-YEAR WILKINSON MICROWAVE ANISOTROPY PROBE (WMAP ) OBSERVATIONS: COSMOLOGICAL INTERPRETATION}",
      collaboration  = "WMAP Collaboration",
      journal        = "Astrophys. J. Suppl.",
      volume         = "192",
      pages          = "18",
      doi            = "10.1088/0067-0049/192/2/18",
      year           = "2011",
      eprint         = "1001.4538",
      archivePrefix  = "arXiv",
      primaryClass   = "astro-ph.CO",
}

@article{Abercrombie:2015wmb,
    author = "Abercrombie, Daniel and others",
    title = "{Dark Matter benchmark models for early LHC Run-2 Searches: Report of the ATLAS/CMS Dark Matter Forum}",
    eprint = "1507.00966",
    archivePrefix = "arXiv",
    primaryClass = "hep-ex",
    reportNumber = "FERMILAB-PUB-15-282-CD",
    doi = "10.1016/j.dark.2019.100371",
    journal = "Phys. Dark Univ.",
    volume = "27",
    pages = "100371",
    year = "2020"
}

@article{Feng_2010,
	doi = {10.1146/annurev-astro-082708-101659},
	%url = {https://doi.org/10.1146%2Fannurev-astro-082708-101659},
	year = 2010,
	month = {aug},
	publisher = {Annual Reviews},
	volume = {48},
	number = {1},
	pages = {495--545},
	author = {Jonathan L. Feng},
        archivePrefix  = "arXiv",
        primaryClass   = "astro-ph.CO",
        eprint = "1003.0904",	
        title = {Dark Matter Candidates from Particle Physics and Methods of Detection},
  
	journal = {Annual Review of Astronomy and Astrophysics}
}

@article{Boveia_2018,
	doi = {10.1146/annurev-nucl-101917-021008},
	%url = {https://doi.org/10.1146%2Fannurev-nucl-101917-021008},
	year = 2018,
	month = {oct},
	publisher = {Annual Reviews},
	volume = {68},
	number = {1},
	pages = {429--459},
        archivePrefix = "arXiv",
        primaryClass = "hep-ex",
        eprint= "1810.12238",
	author = {Antonio Boveia and Caterina Doglioni},  
	title = {Dark Matter Searches at Colliders},
	journal = {Annual Review of Nuclear and Particle Science}
}

@article{Antoniadis:2004se,
      author         = "Antoniadis, Ignatios and Tuckmantel, Marc and Zwirner,
                        Fabio",
      title          = "{Phenomenology of a leptonic goldstino and invisible
                        Higgs boson decays}",
      journal        = "Nucl. Phys. ",
      volume         = "707",
      year           = "2005",
      pages          = "215-232",
      doi            = "10.1016/j.nuclphysb.2004.11.061",
      eprint         = "hep-ph/0410165",
      archivePrefix  = "arXiv",
      primaryClass   = "hep-ph",
      reportNumber   = "CERN-PH-TH-2004-195, ROMA-1390-04",
      SLACcitation   = "%%CITATION = HEP-PH/0410165;%%"
}

@article{ArkaniHamed:1998vp,
      author         = "Arkani-Hamed, Nima and Dimopoulos, Savas and Dvali, G. R.
                        and March-Russell, John",
      title          = "{Neutrino masses from large extra dimensions}",
      booktitle      = "{SUSY 98 Conference Oxford, England, July 11-17, 1998}",
      journal        = "Phys. Rev. D",
      volume         = "65",
      year           = "2001",
      pages          = "024032",
      doi            = "10.1103/PhysRevD.65.024032",
      eprint         = "hep-ph/9811448",
      archivePrefix  = "arXiv",
      primaryClass   = "hep-ph",
      reportNumber   = "SLAC-PUB-8014, SU-ITP-98-64",
      SLACcitation   = "%%CITATION = HEP-PH/9811448;%%"
}

@article{Datta:2004jg,
      author         = "Datta, Anindya and Huitu, Katri and Laamanen, Jari and
                        Mukhopadhyaya, Biswarup",
      title          = "{Linear collider signals of an invisible Higgs boson in theories of large extra dimensions}",
      journal        = "Phys. Rev. D",
      volume         = "70",
      year           = "2004",
      pages          = "075003",
      doi            = "10.1103/PhysRevD.70.075003",
      eprint         = "hep-ph/0404056",
      archivePrefix  = "arXiv",
      primaryClass   = "hep-ph",
      reportNumber   = "HRI-04-04-001, HIP-2004-11-TH, ROME1-1372-2004",
      SLACcitation   = "%%CITATION = HEP-PH/0404056;%%"
}

@article{Shrock:1982kd,
      author         = "Shrock, Robert E. and Suzuki, Mahiko",
      title          = "{Invisible Decays of Higgs Bosons}",
      journal        = "Phys. Lett. B",
      volume         = "110",
      year           = "1982",
      pages          = "250",
      doi            = "10.1016/0370-2693(82)91247-3",
      reportNumber   = "PRINT-82-0199 (UC,BERKELEY)",
      SLACcitation   = "%%CITATION = PHLTA,110B,250;%%"
}

@article{Choudhury:1993hv,
      author         = "Choudhury, Debajyoti and Roy, D. P.",
      title          = "{Signatures of an invisibly decaying Higgs particle at
                        LHC}",
      journal        = "Phys. Lett. B",
      volume         = "322",
      year           = "1994",
      pages          = "368-373",
      doi            = "10.1016/0370-2693(94)91167-3",
      eprint         = "hep-ph/9312347",
      archivePrefix  = "arXiv",
      primaryClass   = "hep-ph",
      reportNumber   = "TIFR-TH-93-64, MPI-PH-93-106",
      SLACcitation   = "%%CITATION = HEP-PH/9312347;%%"
}

@article{Davoudiasl:2004aj,
      author         = "Davoudiasl, Hooman and Han, Tao and Logan, Heather E.",
      title          = "{Discovering an invisibly decaying Higgs boson at hadron colliders}",
      journal        = "Phys. Rev. D",
      volume         = "71",
      year           = "2005",
      pages          = "115007",
      doi            = "10.1103/PhysRevD.71.115007",
      eprint         = "hep-ph/0412269",
      archivePrefix  = "arXiv",
      primaryClass   = "hep-ph",
      reportNumber   = "FERMILAB-PUB-04-371-T, MADPH-04-1408",
      SLACcitation   = "%%CITATION = HEP-PH/0412269;%%"
}

@article{Godbole:2003it,
      author         = "Godbole, R. M. and Guchait, M. and Mazumdar, K. and
                        Moretti, S. and Roy, D. P.",
      title          = "{Search for `invisible' Higgs signals at LHC via
                        associated production with gauge bosons}",
      journal        = "Phys. Lett. B",
      volume         = "571",
      year           = "2003",
      pages          = "184-192",
      doi            = "10.1016/j.physletb.2003.06.066",
      eprint         = "hep-ph/0304137",
      archivePrefix  = "arXiv",
      primaryClass   = "hep-ph",
      reportNumber   = "DESY-03-042, SHEP-03-06, TIFR-EHEP-03-03, TIFR-TH-03-08",
      SLACcitation   = "%%CITATION = HEP-PH/0304137;%%"
}

@article{Ghosh:2012ep,
      author         = "Ghosh, Diptimoy and Godbole, Rohini and Guchait,
                        Monoranjan and Mohan, Kirtimaan and Sengupta, Dipan",
      title          = "{Looking for an Invisible Higgs Signal at the LHC}",
      journal        = "Phys. Lett. B",
      volume         = "725",
      year           = "2013",
      pages          = "344-351",
      doi            = "10.1016/j.physletb.2013.07.042",
      eprint         = "1211.7015",
      archivePrefix  = "arXiv",
      primaryClass   = "hep-ph",
      SLACcitation   = "%%CITATION = ARXIV:1211.7015;%%"
}

@article{Belanger:2013kya,
      author         = "Belanger, G. and Dumont, B. and Ellwanger, U. and Gunion,
                        J. F. and Kraml, S.",
      title          = "{Status of invisible Higgs decays}",
      journal        = "Phys. Lett. B",
      volume         = "723",
      year           = "2013",
      pages          = "340-347",
      doi            = "10.1016/j.physletb.2013.05.024",
      eprint         = "1302.5694",
      archivePrefix  = "arXiv",
      primaryClass   = "hep-ph",
      reportNumber   = "LAPTH-013-13, LPSC-13039, LPT-12-53, UCD-2013-2",
      SLACcitation   = "%%CITATION = ARXIV:1302.5694;%%"
}

@article{Curtin:2013fra,
      author         = "Curtin, David and others",
      title          = "{Exotic decays of the 125 GeV Higgs boson}",
      journal        = "Phys. Rev. D",
      volume         = "90",
      year           = "2014",
      number         = "7",
      pages          = "075004",
      doi            = "10.1103/PhysRevD.90.075004",
      eprint         = "1312.4992",
      archivePrefix  = "arXiv",
      primaryClass   = "hep-ph",
      reportNumber   = "YITP-13-47, PITT-PACC-1314",
      SLACcitation   = "%%CITATION = ARXIV:1312.4992;%%"
}

@article{XENON:2019zpr,
    author = "Aprile, E. and others",
    collaboration = "XENON",
    title = "{Search for Light Dark Matter Interactions Enhanced by the Migdal Effect or Bremsstrahlung in XENON1T}",
    eprint = "1907.12771",
    archivePrefix = "arXiv",
    primaryClass = "hep-ex",
    doi = "10.1103/PhysRevLett.123.241803",
    journal = "Phys. Rev. Lett.",
    volume = "123",
    number = "24",
    pages = "241803",
    year = "2019"
}

@Article{ATLAS:2022vkf,
    author = "{ATLAS Collaboration}",
    collaboration = "ATLAS",
    title = "{A detailed map of Higgs boson interactions by the ATLAS experiment ten years after the discovery}",
    eprint = "2207.00092",
    archivePrefix = "arXiv",
    primaryClass = "hep-ex",
    reportNumber = "CERN-EP-2022-057",
    doi = "10.1038/s41586-022-04893-w",
    journal = "Nature",
    volume = "607",
    number = "7917",
    pages = "52--59",
    year = "2022"
}

@misc{ATLAS:2022ygn,
    author = "{ATLAS Collaboration}",
    collaboration = "ATLAS",
    title = "{Constraints on spin-0 dark matter mediators and invisible Higgs decays using ATLAS 13 TeV $pp$ collision data with two top quarks and missing transverse momentum in the final state}",
    eprint = "2211.05426",
    archivePrefix = "arXiv",
    primaryClass = "hep-ex",
    reportNumber = "CERN-EP-2022-133",
    month = "11",
    year = "2022"
}

@article{Lidnert:2018kmw,
      author         = "Kudashkin, K. and Lindert, J. M. and Melnikov, K. and Wever, C.",
      title          = "{Higgs bosons with large transverse momentum at the LHC}",
      journal        = "Phys. Lett. B",
      volume         = "782",
      year           = "2018",
      pages          = "210",
      doi            = "10.1016/j.physletb.2018.05.009",
      eprint         = "1801.08226",
      archivePrefix  = "arXiv",
      primaryClass   = "hep-ph",
      reportNumber   = "IPPP/18/5, TTP18-03",
      SLACcitation   = "%%CITATION = ARXIV:1801.08226;%%"
}

@article{Djouadi:2018kms,
      author         = "Djouadi, A. and Kalinowski, J. and Mühlleitner, M. and Spira, M.",
      title          = "{HDECAY: Twenty++ years after}",
      journal        = "Comput. Phys. Commun.",
      volume         = "238",
      year           = "2019",
      pages          = "214",
      doi            = "10.1016/j.cpc.2018.12.010",
      eprint         = "1801.09506",
      archivePrefix  = "arXiv",
      primaryClass   = "hep-ph",
      reportNumber   = "CERN-TH-2017-262, LPT-Orsay-18-04, KA-TP-03-2018, PSI-PR-18-02",
      SLACcitation   = "%%CITATION = ARXIV:1801.09506;%%"
}

@article{Cacciari:2018dksz,
      author         = "Cacciari, M. and Dreyer, F. A. and Karlberg, A. and Salam, G. P. and Zanderighi, G.",
      title          = "{Fully Differential Vector-Boson-Fusion Higgs Production at Next-to-Next-to-Leading Order}",
      journal        = "Phys. Rev. Lett.",
      volume         = "115",
      year           = "2015",
      pages          = "082002",
      doi            = "10.1103/PhysRevLett.115.082002",
      eprint         = "1506.02660",
      archivePrefix  = "arXiv",
      primaryClass   = "hep-ph",
      SLACcitation   = "%%CITATION = ARXIV:1506.02660;%%",
       addendum       = "Erratum: \href{https://doi.org/10.1103/PhysRevLett.120.139901}{Phys. Rev. Lett. \textbf{120} (2018) 139901}"
}

@article{deFlorian:2016spz,
      author         = "de Florian, D. and others",
      title          = "{Handbook of LHC Higgs Cross Sections: 4. Deciphering the
                        Nature of the Higgs Sector}",
      collaboration  = "LHC Higgs Cross Section Working Group",
      doi            = "10.23731/CYRM-2017-002",
      year           = "2016",
      eprint         = "1610.07922",
      archivePrefix  = "arXiv",
      primaryClass   = "hep-ph",
      reportNumber   = "FERMILAB-FN-1025-T, CERN-2017-002-M",
      SLACcitation   = "%%CITATION = ARXIV:1610.07922;%%"
}

@article{Dulat:2018rbf,
      author         = "Dulat, Falko and Lazopoulos, Achilleas and Mistlberger,
                        Bernhard",
      title          = "{iHixs 2 -- Inclusive Higgs cross sections}",
      journal        = "Comput. Phys. Commun.",
      volume         = "233",
      year           = "2018",
      pages          = "243-260",
      doi            = "10.1016/j.cpc.2018.06.025",
      eprint         = "1802.00827",
      archivePrefix  = "arXiv",
      primaryClass   = "hep-ph",
      reportNumber   = "CERN-TH-2018-019, SLAC-PUB-17222",
      SLACcitation   = "%%CITATION = ARXIV:1802.00827;%%"
}

@article{Bonetti:2018ukf,
      author         = "Bonetti, Marco and Melnikov, Kirill and Tancredi,
                        Lorenzo",
      title          = "{Higher order corrections to mixed QCD-EW contributions
                        to Higgs boson production in gluon fusion}",
      journal        = "Phys. Rev. D",
      volume         = "97",
      year           = "2018",
      number         = "5",
      pages          = "056017",
      doi            = "10.1103/PhysRevD.97.056017",
      eprint         = "1801.10403",
      archivePrefix  = "arXiv",
      primaryClass   = "hep-ph",
      reportNumber   = "CERN-TH-2018-011, TTP18-004",
      SLACcitation   = "%%CITATION = ARXIV:1801.10403;%%",
      related = "Bonetti:2018ukf-err",
      relatedstring  = "Erratum:",
}

@article{Harlander:2018yio,
      author         = "Harlander, Robert V. and Klappert, Jonas and Liebler,
                        Stefan and Simon, Lukas",
      title          = "{vh@nnlo-v2: new physics in Higgs Strahlung}",
      journal        = "JHEP",
      volume         = "05",
      year           = "2018",
      pages          = "089",
      doi            = "10.1007/JHEP05(2018)089",
      eprint         = "1802.04817",
      archivePrefix  = "arXiv",
      primaryClass   = "hep-ph",
      reportNumber   = "KA-TP-01-2018, TTK-17-47",
      SLACcitation   = "%%CITATION = ARXIV:1802.04817;%%"
}

@Article{Read:2002hq,
    Author = {Read, Alexander L.},
    Journal = {J. Phys. G},
    Pages = {2693},
    doi = "10.1088/0954-3899/28/10/313",
    Title = "{Presentation of search results: the \(CL_S\) technique}",
    Volume = {28},
    Year = {2002}
}

@Article{Cowan:2010js,
   Author = {Cowan, Glen and Cranmer, Kyle and Gross, Eilam and Vitells, Ofer},
   Title = {{Asymptotic formulae for likelihood-based tests of new physics}},
   Journal = {Eur. Phys. J. C},
   Volume = {71},
   Year = {2011},
   Pages = {1554},
   doi = "10.1140/epjc/s10052-011-1554-0",
   Eprint = "1007.1727",
   Archiveprefix = {arXiv},
   Primaryclass = {physics.data-an},
   related = "Cowan:2010js-err",
   relatedstring  = "Erratum:",
}

@Booklet{RooFit,
      author         = "Verkerke, Wouter and Kirkby, David",
      title          = "{The RooFit toolkit for data modeling}",
      year           = "2003",
      eprint         = "physics/0306116",
      archivePrefix  = "arXiv",
      primaryClass   = "physics.data-an",
}

@Article{PERF-2007-01,
    author         = "{ATLAS Collaboration}",
    title          = "{The ATLAS Experiment at the CERN Large Hadron Collider}",
    journal        = "JINST",
    volume         = "3",
    year           = "2008",
    pages          = "S08003",
    doi            = "10.1088/1748-0221/3/08/S08003",
    primaryClass   = "hep-ex",
}

@Article{HIGG-2012-27,
    author         = "{ATLAS Collaboration}",
    title          = "{Observation of a new particle in the search for the Standard Model Higgs boson with the ATLAS detector at the LHC}",
    journal        = "Phys. Lett. B",
    volume         = "716",
    year           = "2012",
    pages          = "1",
    doi            = "10.1016/j.physletb.2012.08.020",
    reportNumber   = "CERN-PH-EP-2012-218",
    eprint         = "1207.7214",
    archivePrefix  = "arXiv",
    primaryClass   = "hep-ex",
}

@Article{HIGG-2013-03,
    author         = "{ATLAS Collaboration}",
    title          = "{Search for Invisible Decays of a Higgs Boson Produced in Association with a \(Z\) Boson in ATLAS}",
    journal        = "Phys. Rev. Lett.",
    volume         = "112",
    year           = "2014",
    pages          = "201802",
    doi            = "10.1103/PhysRevLett.112.201802",
    reportNumber   = "CERN-PH-EP-2013-210",
    eprint         = "1402.3244",
    archivePrefix  = "arXiv",
    primaryClass   = "hep-ex",
}

@Article{HIGG-2013-16,
    author         = "{ATLAS Collaboration}",
    title          = "{Search for invisible decays of a Higgs boson using vector-boson fusion in \(pp\) collisions at \(\sqrt{s} = 8\,\text{TeV}\) with the ATLAS detector}",
    journal        = "JHEP",
    volume         = "01",
    year           = "2016",
    pages          = "172",
    doi            = "10.1007/JHEP01(2016)172",
    reportNumber   = "CERN-PH-EP-2015-186",
    eprint         = "1508.07869",
    archivePrefix  = "arXiv",
    primaryClass   = "hep-ex",
}

@Article{HIGG-2014-07,
    author         = "{ATLAS Collaboration}",
    title          = "{Search for invisible decays of the Higgs boson produced in association with a hadronically decaying vector boson in \(pp\) collisions at \(\sqrt{s} = 8\,\text{TeV}\) with the ATLAS detector}",
    journal        = "Eur. Phys. J. C",
    volume         = "75",
    year           = "2015",
    pages          = "337",
    doi            = "10.1140/epjc/s10052-015-3551-1",
    reportNumber   = "CERN-PH-EP-2015-061",
    eprint         = "1504.04324",
    archivePrefix  = "arXiv",
    primaryClass   = "hep-ex",
}

@Article{HIGG-2015-03,
    author         = "{ATLAS Collaboration}",
    title          = "{Constraints on new phenomena via Higgs boson couplings and invisible decays with the ATLAS detector}",
    journal        = "JHEP",
    volume         = "11",
    year           = "2015",
    pages          = "206",
    doi            = "10.1007/JHEP11(2015)206",
    reportNumber   = "CERN-PH-EP-2015-191",
    eprint         = "1509.00672",
    archivePrefix  = "arXiv",
    primaryClass   = "hep-ex",
}

@Article{EXOT-2018-06,
    author         = "{ATLAS Collaboration}",
    title          = "{Search for new phenomena in events with an energetic jet and missing transverse momentum in \(pp\) collisions at \(\sqrt{s} = 13\,\text{TeV}\) with the ATLAS detector}",
    journal        = "Phys. Rev. D",
    volume         = "103",
    year           = "2021",
    pages          = "112006",
    doi            = "10.1103/PhysRevD.103.112006",
    reportNumber   = "CERN-EP-2020-238",
    eprint         = "2102.10874",
    archivePrefix  = "arXiv",
    primaryClass   = "hep-ex",
}

@Article{FTAG-2018-01,
    author         = "{ATLAS Collaboration}",
    title          = "{ATLAS \(b\)-jet identification performance and efficiency measurement with \(t\bar{t}\) events in \(pp\) collisions at \(\sqrt{s} = 13\,\text{TeV}\)}",
    journal        = "Eur. Phys. J. C",
    volume         = "79",
    year           = "2019",
    pages          = "970",
    doi            = "10.1140/epjc/s10052-019-7450-8",
    reportNumber   = "CERN-EP-2019-132",
    eprint         = "1907.05120",
    archivePrefix  = "arXiv",
    primaryClass   = "hep-ex",
}

@Article{HIGG-2018-26,
    author         = "{ATLAS Collaboration}",
    title          = "{Search for associated production of a \(Z\) boson with an invisibly decaying Higgs boson or dark matter candidates at \(\sqrt{s} = 13\,\text{TeV}\) with the ATLAS detector}",
    journal        = "Phys. Lett. B",
    volume         = "829",
    year           = "2021",
    pages          = "137066",
    doi            = "10.1016/j.physletb.2022.137066",
    reportNumber   = "CERN-EP-2021-204",
    eprint         = "2111.08372",
    archivePrefix  = "arXiv",
    primaryClass   = "hep-ex",
}

@Article{HIGG-2018-54,
    author         = "{ATLAS Collaboration}",
    title          = "{Combination of Searches for Invisible Higgs Boson Decays with the ATLAS Experiment}",
    journal        = "Phys. Rev. Lett.",
    volume         = "122",
    year           = "2019",
    pages          = "231801",
    doi            = "10.1103/PhysRevLett.122.231801",
    reportNumber   = "CERN-EP-2019-046",
    eprint         = "1904.05105",
    archivePrefix  = "arXiv",
    primaryClass   = "hep-ex",
}

@Article{SUSY-2018-07,
    author         = "{ATLAS Collaboration}",
    title          = "{Search for new phenomena with top quark pairs in final states with one lepton, jets, and missing transverse momentum in \(pp\) collisions at \(\sqrt{s} = 13\,\text{TeV}\) with the ATLAS detector}",
    journal        = "JHEP",
    volume         = "04",
    year           = "2020",
    pages          = "174",
    doi            = "10.1007/JHEP04(2021)174",
    reportNumber   = "CERN-EP-2020-177",
    eprint         = "2012.03799",
    archivePrefix  = "arXiv",
    primaryClass   = "hep-ex",
}

@Article{SUSY-2018-08,
    author         = "{ATLAS Collaboration}",
    title          = "{Search for new phenomena in events with two opposite-charge leptons, jets and missing transverse momentum in \(pp\) collisions at \(\sqrt{s} = 13\,\text{TeV}\) with the ATLAS detector}",
    journal        = "JHEP",
    volume         = "04",
    year           = "2021",
    pages          = "165",
    doi            = "10.1007/JHEP04(2021)165",
    reportNumber   = "CERN-EP-2020-243",
    eprint         = "2102.01444",
    archivePrefix  = "arXiv",
    primaryClass   = "hep-ex",
}
 
\clearpage
 
\begin{flushleft}
\hypersetup{urlcolor=black}
{\Large The ATLAS Collaboration}

\bigskip

\AtlasOrcid[0000-0002-6665-4934]{G.~Aad}$^\textrm{\scriptsize 102}$,
\AtlasOrcid[0000-0002-5888-2734]{B.~Abbott}$^\textrm{\scriptsize 120}$,
\AtlasOrcid[0000-0002-1002-1652]{K.~Abeling}$^\textrm{\scriptsize 55}$,
\AtlasOrcid[0000-0002-8496-9294]{S.H.~Abidi}$^\textrm{\scriptsize 29}$,
\AtlasOrcid[0000-0002-9987-2292]{A.~Aboulhorma}$^\textrm{\scriptsize 35e}$,
\AtlasOrcid[0000-0001-5329-6640]{H.~Abramowicz}$^\textrm{\scriptsize 151}$,
\AtlasOrcid[0000-0002-1599-2896]{H.~Abreu}$^\textrm{\scriptsize 150}$,
\AtlasOrcid[0000-0003-0403-3697]{Y.~Abulaiti}$^\textrm{\scriptsize 117}$,
\AtlasOrcid[0000-0003-0762-7204]{A.C.~Abusleme~Hoffman}$^\textrm{\scriptsize 137a}$,
\AtlasOrcid[0000-0002-8588-9157]{B.S.~Acharya}$^\textrm{\scriptsize 69a,69b,n}$
% ...
\bigskip
\\

% ...
$^{29}$Physics Department, Brookhaven National Laboratory, Upton NY; United States of America.\\
% ...
$^{35}$$^{(a)}$Facult\'e des Sciences Ain Chock, R\'eseau Universitaire de Physique des Hautes Energies - Universit\'e Hassan II, Casablanca;$^{(b)}$Facult\'{e} des Sciences, Universit\'{e} Ibn-Tofail, K\'{e}nitra;$^{(c)}$Facult\'e des Sciences Semlalia, Universit\'e Cadi Ayyad, LPHEA-Marrakech;$^{(d)}$LPMR, Facult\'e des Sciences, Universit\'e Mohamed Premier, Oujda;$^{(e)}$Facult\'e des sciences, Universit\'e Mohammed V, Rabat;$^{(f)}$Institute of Applied Physics, Mohammed VI Polytechnic University, Ben Guerir; Morocco.\\
% ...
$^{55}$II. Physikalisches Institut, Georg-August-Universit\"{a}t G\"ottingen, G\"ottingen; Germany.\\
% ...
$^{69}$$^{(a)}$INFN Gruppo Collegato di Udine, Sezione di Trieste, Udine;$^{(b)}$ICTP, Trieste;$^{(c)}$Dipartimento Politecnico di Ingegneria e Architettura, Universit\`a di Udine, Udine; Italy.\\
% ...
$^{102}$CPPM, Aix-Marseille Universit\'e, CNRS/IN2P3, Marseille; France.\\
% ...
$^{117}$Department of Physics, New York University, New York NY; United States of America.\\
% ...
$^{120}$Homer L. Dodge Department of Physics and Astronomy, University of Oklahoma, Norman OK; United States of America.\\
% ...
$^{137}$$^{(a)}$Departamento de F\'isica, Pontificia Universidad Cat\'olica de Chile, Santiago;$^{(b)}$Millennium Institute for Subatomic physics at high energy frontier (SAPHIR), Santiago;$^{(c)}$Instituto de Investigaci\'on Multidisciplinario en Ciencia y Tecnolog\'ia, y Departamento de F\'isica, Universidad de La Serena;$^{(d)}$Universidad Andres Bello, Department of Physics, Santiago;$^{(e)}$Instituto de Alta Investigaci\'on, Universidad de Tarapac\'a, Arica;$^{(f)}$Departamento de F\'isica, Universidad T\'ecnica Federico Santa Mar\'ia, Valpara\'iso; Chile.\\
% ...
$^{150}$Department of Physics, Technion, Israel Institute of Technology, Haifa; Israel.\\
$^{151}$Raymond and Beverly Sackler School of Physics and Astronomy, Tel Aviv University, Tel Aviv; Israel.\\
% ...
$^{a}$ Also Affiliated with an institute covered by a cooperation agreement with CERN.\\
$^{b}$ Also at Borough of Manhattan Community College, City University of New York, New York NY; United States of America.\\
$^{c}$ Also at Bruno Kessler Foundation, Trento; Italy.\\
$^{d}$ Also at Center for High Energy Physics, Peking University; China.\\
$^{e}$ Also at Center for Interdisciplinary Research and Innovation (CIRI-AUTH), Thessaloniki ; Greece.\\
$^{f}$ Also at Centro Studi e Ricerche Enrico Fermi; Italy.\\
% ...
$^{n}$ Also at Department of Physics, King's College London, London; United Kingdom.\\
% ...

\end{flushleft}


\end{document}